The final published version is available at https://onlinelibrary.wiley.com/doi/10.1002/ima.22933
Code is available at https://github.com/xia-xx-cv/LANet-DR/

# Lesion-aware Network for Diabetic Retinopathy Diagnosis

Xue Xia[1] | Kun Zhan[1] | Yuming Fang[1] | Wenhui Jiang[1] | Fei Shen[2]

[1]Org Division, Jiangxi University of Finance and Economics, Jiangxi, China
[2]Org Division, Sany Heavy Industry Co. Ltd., Beijing, China

**Abstract**

Deep learning brought boosts to auto Diabetic Retinopathy (DR) diagnosis, thus greatly help ophthalmologists for early disease detection, which contribute to prevent disease deterioration that may eventually lead to blindness. It has been proved that Convolutional Neural Network(CNN)-aided lesion identifying or segmentation benefits auto DR screening. The key of fine-grained lesion tasks mainly lie in: 1) extracting discriminative features being both sensitive to tiny lesion areas and robust to DR-irrelevant interference, 2) learning lesion features from images with extremely imbalanced data distribution. To this end, we propose CNN-based DR diagnosis network with attention mechanism involved, termed Lesion-Aware Network (LANet), to better capture lesion information from imbalanced data. Specifically, we design the Lesion-Aware Module (LAM) to capture noise-like lesion areas across deeper layers, and the Feature-Preserve Module (FPM) to assist shallow-to-deep feature fusion. Afterwards, the LANet is constructed by embedding LAM and FPM into the CNN decoders for DR-related information utilization. The LANet is then further extended to a DR screening network by simply adding a classification layer. Through experiments on three fundus datasets with pixel-level annotations, our method outperforms the mainstream methods with an AUC of 0.967 in DR screening, and increases the mAP by 7.6%, 2.1% and 1.2% in lesion segmentation on three datasets. Besides, the ablation study validates the effectiveness of the proposed sub-modules.

## 1 | INTRODUCTION

Diabetic Retinopathy (DR) is a kind of retinal complication, which may lead to vision loss or blindness if left untreated [1], caused by long term effects of diabetes. Generally, observable symptoms may not include in the early or more reversible stages of DR, which means naked eye-visible symptoms are always brought on by progressive lesions in patients with poorly controlled disease. It is reported that by 2019, the number of people diagnosed with diabetes has grown to 463 million [2], and it is estimated that 4.2 million deaths attributable to diabetes [3]. Among all diabetic patients, about one-third suffer from capillaries impairment [4], which can be observed through funduscopy [5] but not naked eyes. Conventional DR-related lesions include microaneurysm (MA), hemorrhages (HE), hard exudates (EX), and soft exudates (SE), as illustrated in Fig. 1 The tiny lesions bring difficulties to ophthalmologists in diagnosis since they appear to be indistinguishable in fundus images at early stages. Fortunately, machine learning based eye disease screening can detect early changes in and around the retina blood vessels [6], yet there are still some obstacles. For instance, some MA and EX presenting noise-like appearances are prone to be over-smoothed by image pre-processing. What's more, the overall structures and color distribution among different fundus images are similar, which leads to low intra-class variance and will further affects the diagnosis. The Convolutional Neural Networks (CNNs) are able to capture semantic information for better DR identification by the hierarchical structure. Therefore, detecting DR-related lesions in early times by deep learning algorithms contributes a lot to vision deterioration or even blind prevention.

[0]**Abbreviations:** LANet, leision-aware network; DR, diabetic retinopathy; LAM, lesion-aware module; FPM, feature preserve module

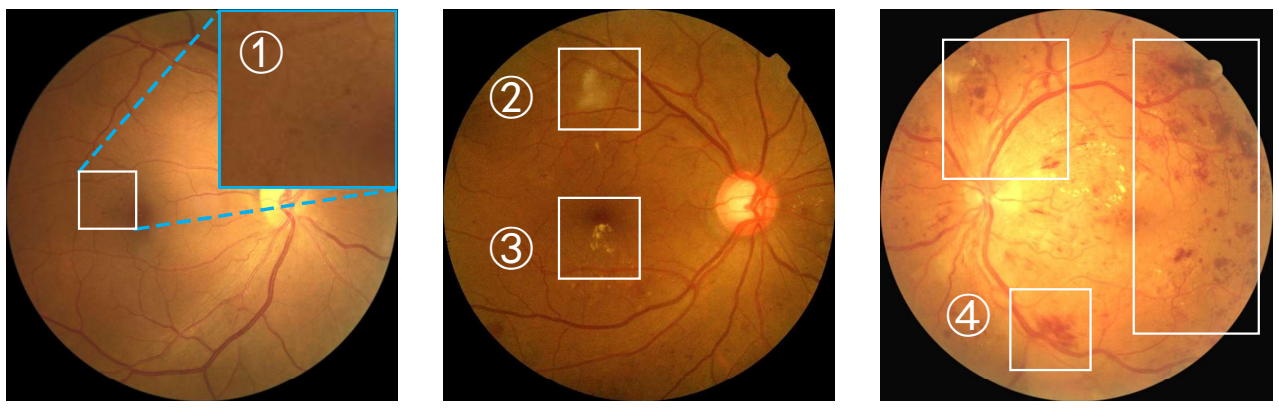

(a) (b) (c)

**FIGURE 1** Examples of DR-related lesions: ①MA, ②SE, ③EX and ④HE.

To meet the demands of DR screening, accurate classification, detection, and segmentation algorithms serve as crucial parts for performing automatic disease grading and lesion identification. Since accurate lesion segmentation can greatly help disease grading [7], we mainly focus on lesion segmentation. However, DR lesion segmentation and screening generally face three main challenges. First, although microaneurysms are the earliest clinically visible symptoms of DR, they occupy extremely small areas compared to anatomical structures of retina, which will easily bring false negative. Second, interference being irrelevant to DR is sometimes prone to be amplified through the convolution and non-linear operations, thus will eventually affect the final DR grading results. Third, the distribution of both pixel-level and image-level DR data are extremely imbalanced, in which lesion pixels account for a very small part. Imbalanced data distribution often leads the model to bias towards the category with more samples, thus will greatly suppress the generalization ability of the model.

To address the above three problems, we aim at proposing a CNN-based network for DR-related lesion segmentation, which is also flexible to be adapted into DR screening task. The key lies in : 1) designing specific modules to assist feature extractor in obtaining discriminative features being both sensitive to tiny lesion areas and robust to DR-irrelevant interference, 2) guiding the model to learn lesion features from data with extremely imbalanced distribution progressively rather than all at once. The former can be implemented by involving attention mechanism into deep networks, while the latter may rely on multi-task learning technique.

Based on the above analysis, we propose the Lesion-Aware Network (LANet) based on attention mechanism for pixel-level DR lesion segmentation and image-level DR screening. The contributions of our work can be summarized as follows:

- We design a Lesion-Aware Attention module (LAM) and a Feature-Preserve Module (FPM) to capture and represent DR-related information. Both modules can be embedded into mainstream backbones for either lesion segmentation or DR screening task. The LAM captures tiny lesion areas across deeper layers, and the FPM assists shallow-to-deep feature fusion for more accurate disease recognition.
- By aggregating the proposed LAM and FPM modules into the encoder-decoder architecture, we construct LANet, which can perform lesion segmentation even under imbalanced data, since the DR-related information is utilized gradually across layers. Moreover, we further extend LANet to a DR Screening Network, termed LASNet, in a simple yet effective way.
- We validate the ability of our net in dealing with imbalanced fundus data and in locating lesions through ablation study. And our method establishes some new SOTAs in lesion identification and DR screening.

## 2 | RELATED WORK

As mentioned above, lesion segmentation matters in medical image analysis [8] and is especially crucial for fundus image-based DR diagnosis. According to the International Classification of Diabetic Retinopathy (ICDR) [9], DR can be broadly divided into two stages, *i.e.*, non-proliferative DR (NPDR) and proliferative DR (PDR), in which the former contains mild DR, moderate DR and severe DR. To coarsely figure out whether a fundus suffers from non-proliferative DR or not is termed as *screening*, while to classify a fundus into a specific severity scale is termed as *grading*. Lesion segmentation or identification can support both screening and grading. We focus on segmentation and screening.

## 2.1 | Attention-based Lesion Segmentation.

Wang *et al.* utilized CNN to diagnose DR and proved through activation maps that the network could focus on specific areas in disease classification [10]. However, they did not consider the importance of specific lesions in DR diagnosis. Jiang *et al.* modeled lesion detection task into a multi-label image classification problem, thus a conventional classification netowrk was adopted. The gradient-weighted class activation mapping was involved based on the last convolutional features and the class scores. The acquired weight map and guided-propagation map intrinsically worked as either classification guidance or attention [11]. These works prove that attention helps lesion detection, which will further provide assistance to DR identification or grading.

Fu *et al.* [12,13] firstly proposed channel attention and spatial attention to combine local features with their related global information. By applying the two attention modules in parallel and weighted summing their results, a dual attention network was conducted to selective feature aggregation, which benefited segmentation precision. Based on this, He *et al.* [14] involved the spatial and channel attentions in sequence as a class-agnostic global feature extractor followed by a novel category attention block(CAB) for DR grading. The CAB calculates class-aware and cross-class attentions from randomly dropped feature relations and full features respectively to deal with imbalanced distributed data. Playout *et al.* proposed a UNet[15]-based network for lesion detection and segmentation [16], but the network only deals with red and bright lesions. Gao *et al.* proposed the CAR-Net[17], which extracts local and global features from both whole images and image patches, and integrates multilevel context features by the proposed attention refinement module. These works verify that attention mechanism helps to support pixel-wise segmentation and further offer finer-grained information to DR screening and grading.

## 2.2 | Multi-task Learning for DR Disgnosis.

Based on the above, some works adopted multi-task learning for DR grading, in which lesion segmentation was always involved as the support [8,18]. Athalye *et al.* focused on DR classification through detecting exudates based on the blood vessel and optic disc segmentation model and identifying microaneurysms through wavelet model [19]. Lin *et al.* explored an anti-noise method that calculated lesion type for each feature position and clustered them to obtain lesion centers, according to which the impact of noisy non-lesion samples were down-weighted for lesion detection. Then the lesion map and original fundus were combined through attention fusion for DR grading [20]. Wang *et al.* leveraged individual sub-networks to identify diseases affecting optic-disc, macula and entire retina. Hence, semantic multitask learning was adopted to explore disease signs for different fundus component regions [21]. This work is able to recognize 36 retinal diseases, and it proves the effectiveness of multitask learning. However, it's more efficient on diseases with obvious regionally appearances due to the shared feature extractor. Therefore, designing network aiming on a specific disease matters.

Yang *et al.* proposed a two-stage network that extracted a heatmap indicating lesions in the first stage, and then applied the heatmap as an imbalanced attention for classification in the second-stage [22]. Wang *et al.* regarded DR grading as the main task and involved both image super-resolution and lesion segmentation as auxiliary tasks [8]. To ensure accuracy, feature selected based on IoU between feature maps and lesion maps and gradient-weighted feature combination were proposed. However, these works depended on sub-networks, which brought difficulties to model training. Thus, we aim to propose a simpler framework for simultaneous DR lesion segmentation and screening through module sharing.

## 2.3 | Learning with Imbalanced Data.

One thing that hinders lesion identification and lesion feature learning is the lack of annotation, and another is the imbalanced distribution of lesion pixels and non-ones. According to the two obstacles, lesion segmentation or lesion-related diagnosis can be viewed as tasks with scarce samples. Generally used strategies for model learning with limited fundus samples are data augmentation, weakly supervised learning, generative adversarial network (GAN) and so on. Different from popular solution to scarce sample learning, Hassan *et al.* proposed a Bayesian integrated deep model for retinopathy screening in OCT and fundus imagery. For DR diagnosis, Zhou *et al.* involved generative adversarial network to generate fundus images with manipulable grading and lesion information in high resolution [23]. This is an alternative to data augmentation and is implemented through learning an adaptive grading vector modeled in a latent space. Nevertheless, generating fundus with abundant vessels and versatile lesions is complicated in computation and time costing. As alternatives, Foo *et al.* adopted semi-supervised learning for lesion segmentation, in which grading labels were involved to correct results of healthy samples [18]. Gondal *et al.* employed class activation mapping compute lesion locations using only image-level labels, which can be regard as a weakly supervised model [24]. Except for scarce data, extremely imbalanced data distribution remains a problem bringing bias to classifiers. Wong *et al.* improved focal loss[25] by considering object relative sizes for segmenting objects in highly unbalanced sizes [26]. This validates the effectiveness of adapting loss function in medical image segmentation.

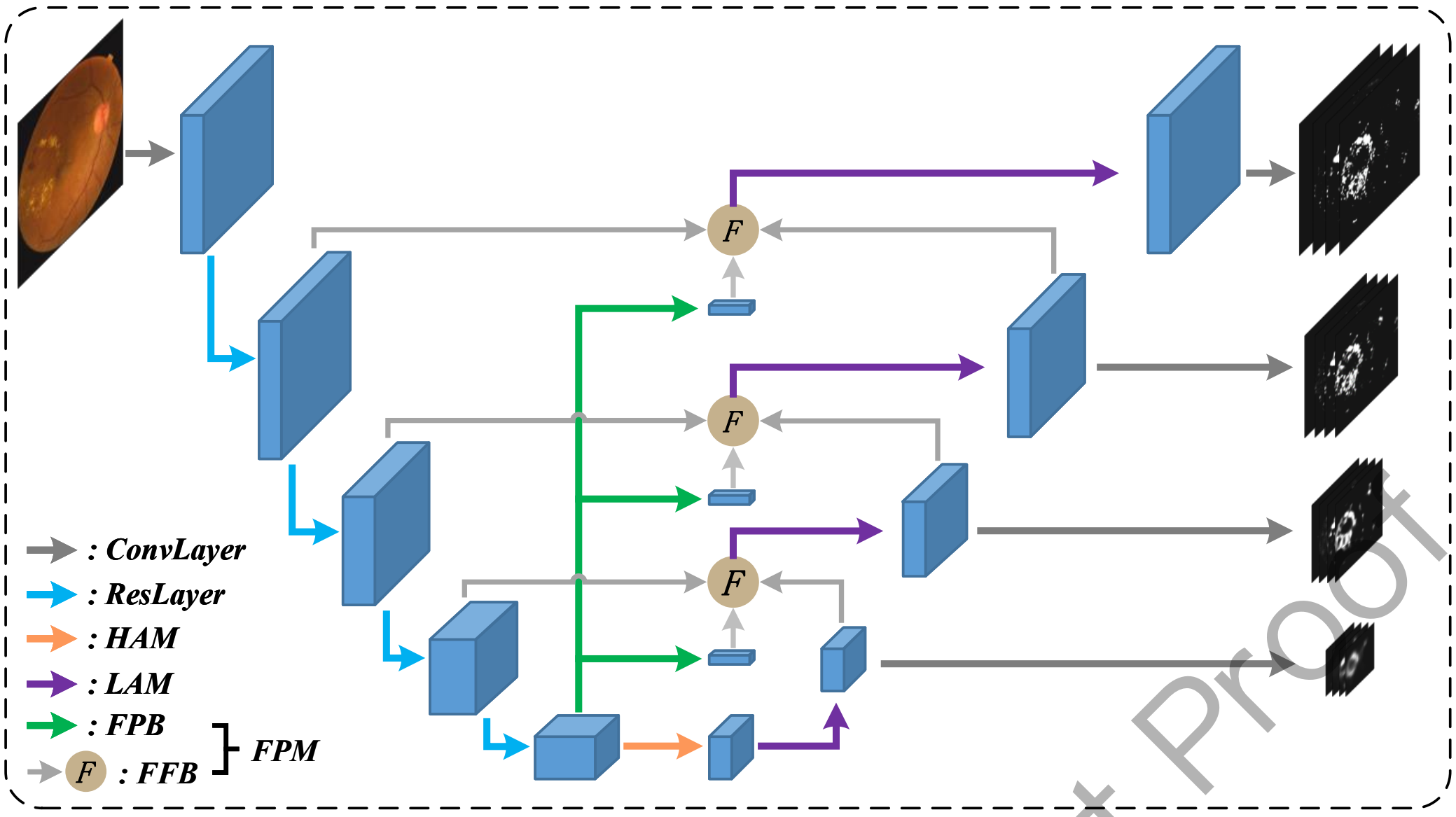

**FIGURE 2** Pipeline of the lesion-aware network for DR segmentation.

In conclusion, most existing methods leverage attention mechanism for lesion localization, involve multi-task for grading assisting, and design specific modules or loss functions for imbalanced data or limited samples learning. However, 1) attention modules are always adopted instead of being designed, 2) sub-networks with heavy weights are usually inevitable, and 3) carefully learning strategy or tricks are needed. In this work, we aim at presenting a DR screening method that relies on lesion segmentation. Specifically, we mainly focus on developing lesion segmentation network to assist DR screening. To reduce training complexity, the segmentation and screening task is supposed to share the same network. To deal with imbalanced data distribution, the network is designed to progressively locate lesion areas through attention structure across multi-layers. Besides, our DR screening only consider NPDR since patients will experience obviously severe visual impairment in the proliferative stage.

## 3 | METHODOLOGY

As shown in Fig. 2, the proposed LANet based on an encoder-decoder structure mainly consists of a Lesion-Aware Module (LAM) and a Feature-Preserve Module (FPM). The former was designed to progressively guide the network capture small lesion-related areas through attention. While the latter contains a Feature-Preserve Block (FPB) that maintains lesion information during feature forwarding and a Feature Fusion Block (FFB) for involving global disease-related information in feature fusion.

### 3.1 | Lesion-Aware Network (LANet)

In our LANet, ResNet-50 [27] is adopted as the encoder for base feature extraction, and the last encoding layer is followed by an attention-based dimension reduction layer to work as the *Head* of our network. While the decoder is stacked by LAMs and FPMs for accurate lesion and disease identification.

Being different from existing short connections between features of encoding stages and their corresponding decoding stages, we design the FPM to fuse features with larger receptive field from FPB, high-level features from a former decoding layer and low-level features from encoder. The fused feature of a decoding layer containing global, local and multi-level information is fed to the next decoding layer after an LAM, which explores lesion areas through attention mechanism.

The blue boxes in Fig. 2 represent feature maps of different sizes. The bold lines with arrows indicate specific operations and data flow, as shown in the legend at the bottom right of Fig. 2. The gray circle marked with "$F$" and pointed by arrows stands for the proposed FPM. Every decoding layer outputs four lesion maps indicating HE, MA, EX and SE respectively. The features in decoding layers are computed through Eq. (1):

$$\begin{cases} \boldsymbol{x}_{\text{dec}}^{i} = f_{\text{LAM}}^{i}(f_{\text{FFB}}^{i}(\boldsymbol{x}_{\text{enc}}^{4-i}, f_{\text{FPB}}^{i}(\boldsymbol{x}_{\text{enc}}^{4}), \boldsymbol{x}_{\text{dec}}^{i-1})) & i > 0 \\ \boldsymbol{x}_{\text{dec}}^{i} = f_{\text{LAM}}^{i}(f_{\text{HAM}}(\boldsymbol{x}_{\text{enc}}^{4}) & i = 0 \end{cases} \tag{1}$$

where $\boldsymbol{x}_{\text{dec}}^{i}$ stands for features from the $i$-th decoding layer. $\boldsymbol{x}_{\text{enc}}^{4}$ is the feature of the last (i.e., the 4-th in Fig. 2) encoding layer. $f_{\text{LAM}}^{i}$, $f_{\text{FFB}}^{i}$ and $f_{\text{FPB}}^{i}$ represent the $i$-th LAM, FFB and FPB, respectively. Although different FPBs accept the same input $\boldsymbol{x}_{\text{enc}}^{4}$, they do not share weights. The $f_{\text{HAM}}$ is a self-attention layer implemented by convolution that avoids expensive matrix multiplication. The $\boldsymbol{x}_{\text{enc}}^{4}$ and $f_{\text{HAM}}$ can be viewed as a simple *head* of our network, which is denoted as HAM (Head Attention Module).

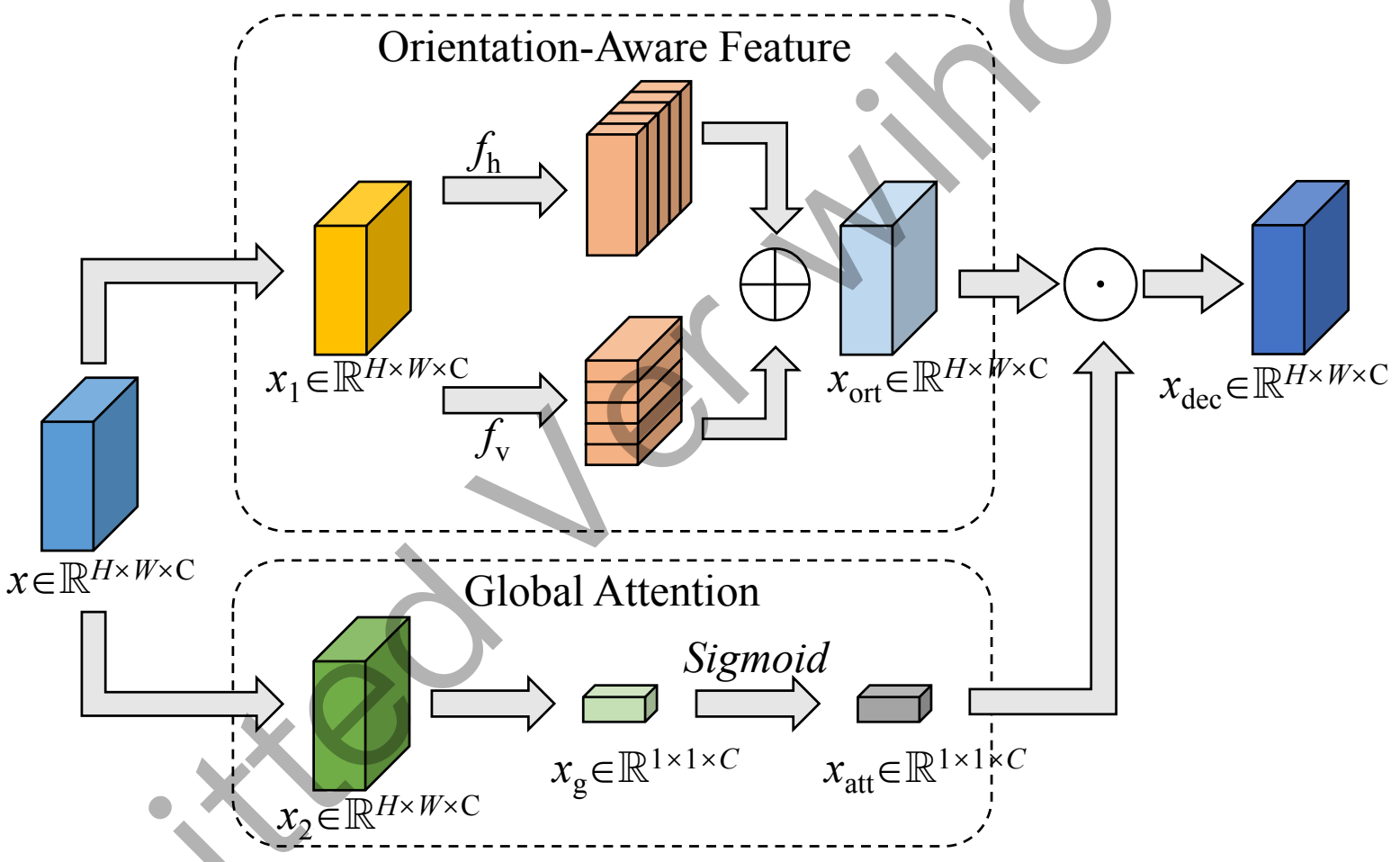

**FIGURE 3** Lesion-aware module that involves orientation-aware features and global attention.

**Lesion-Aware Module (LAM).** The LAM is designed to extract lesion information by orientation-aware features and self-attention mechanism. In Fig. 3, the gray arrows are operations, including convolution, batch normalization or non-linear activation. The plus and multiplication signs in circles stand for element-wise computations. Accordingly, the data flow can be represent by Eq. (2), in which $\boldsymbol{x}_{\text{att}}$ is the attention map computed through Eq. (3).

$\boldsymbol{x}_1$ and $\boldsymbol{x}_2$ are features computed from $\boldsymbol{x}$, the input of the LAM. All $f(\cdot)$ operations in our work are implemented by convolutional layers. $f_{\text{h}}$ and $f_{\text{v}}$ stand for horizontal and vertical spatial convolutions, which are used to describe orientation information.

$$\begin{cases} \boldsymbol{x}_{\text{ort}} = f_{\text{ort}}(f_{\text{h}}(\boldsymbol{x}_1) + f_{\text{v}}(\boldsymbol{x}_1)) \\ \boldsymbol{x}_{\text{dec}}(i,j,c) = \boldsymbol{x}_{\text{ort}}(i,j,c) \times \boldsymbol{x}_{\text{att}}(c) \end{cases} \tag{2}$$

$\boldsymbol{x}_{\text{ort}}$ stands for the orientation-aware feature computed based on information in different orientations. $i, j$ and $c$ stand for the indices along different dimensions of feature maps. In implementation, the lesion-aware feature $\boldsymbol{x}_{\text{dec}}$ is the Hadamard product of orientation feature $\boldsymbol{x}_{\text{ort}}$ and global attention map $\boldsymbol{x}_{\text{att}}$. The latter should be spatially repeated to the same order as $\boldsymbol{x}_{\text{ort}}$ first.

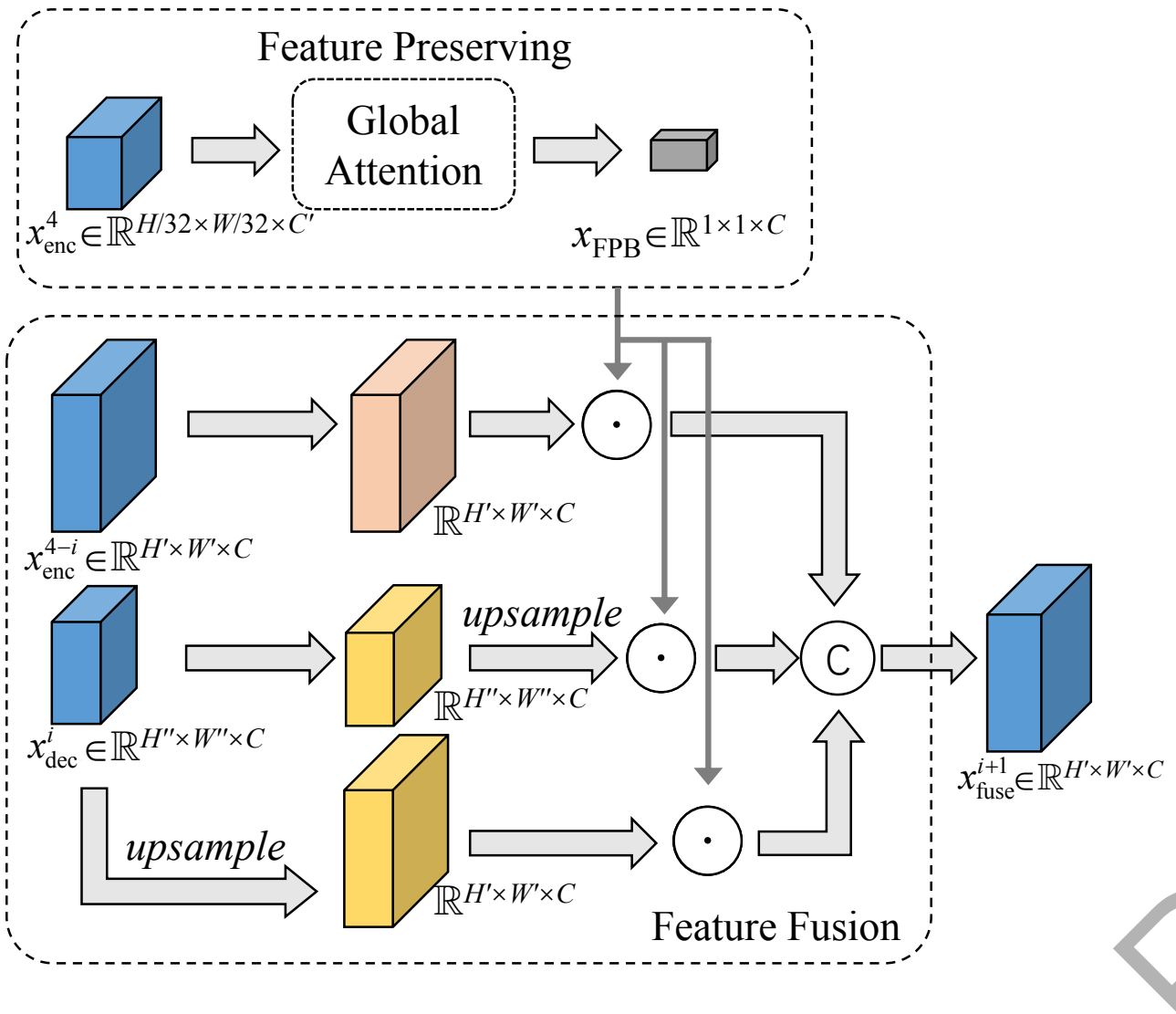

**FIGURE 4** Feature-preserve module that fuses multiple features and preserve lesion-related information.

$$\begin{cases} \boldsymbol{x}_{\text{g}} = f_{\text{conv}}\left(\frac{1}{H\times W}\sum_{i}^{H}\sum_{j}^{W}\boldsymbol{x}_2(i,j,c)\right) \\ \boldsymbol{x}_{\text{att}} = \frac{e^{-\boldsymbol{x}_{\text{g}}}}{1+e^{-\boldsymbol{x}_{\text{g}}}} \end{cases} \tag{3}$$

Eq. (3) corresponds to the Global Attention block in Fig. 3 , where $H$ and $W$ represent the height and width of feature maps, and $e$ denotes the natural exponential function. $f_{\text{conv}}$ represents convolutional layer(s), and $\boldsymbol{x}_{\text{g}}$ and $\boldsymbol{x}_{\text{att}}$ are the global feature and the corresponding global attention map of the input feature $\boldsymbol{x}$. Thus, $\boldsymbol{x}_{\text{att}}$ is a channel attention map that contains global attention, since spatial information of a whole feature map is squeezed while cross-channel information is preserved.

As demonstrated above, our LAM is intrinsically a convolution-based self-attention module, which involves a channel attention modification as the attention map extractor. The main differences between LAM and existing self-attention computations are two folds: 1) The channel number remains the same during global attention computation. 2) Rather than applying attention to self-feature $\boldsymbol{x}_{\text{dec}}$, the computed global attention map $\boldsymbol{x}_{\text{att}}$ is leveraged to orientation-aware feature $\boldsymbol{x}_{\text{ort}}$ in another branch. As a result, our LAM works as a lesion guidance and presents a cross-branch-attention that converts orientation-aware features into lesion-aware features. Thus, the output $\boldsymbol{x}_{\text{out}} = f_{\text{out}}(\boldsymbol{x}_{\text{dec}})$, where $f_{\text{out}}(\cdot)$ represents the convolutional layer that outputs lesion maps. $\boldsymbol{x}_{\text{out}}$ is in size of $H\times W\times 4$, and the index $m = \{1,2,3,4\}$.

**Feature-Preserve Module (FPM).** Simply concatenating feature maps from different layers may limited the representation ability since these features hold different semantics[28]. Therefore, We propose this module to preserve two kinds of features, one is multi-scale feature obtained by the encoder, the other is multi-layer features from shallow-to-deep layers. The former is implemented by Feature-Preserve Block (FPB) and the latter by Feature Fusion Block (FFB).

Inspired by existing pyramid-based multi-scale feature extracting modules like ASPP [29], we propose the FPB to pass $\boldsymbol{x}_{\text{enc}}^4$, global features with the largest receptive field, to decoded features with different resolutions as the information preserver. In addition, to preserve shape-aware features in shallower layers of the network, lesion-aware features from LAM and semantic features in deeper layers, we propose the FFB that progressively aggregate these features.

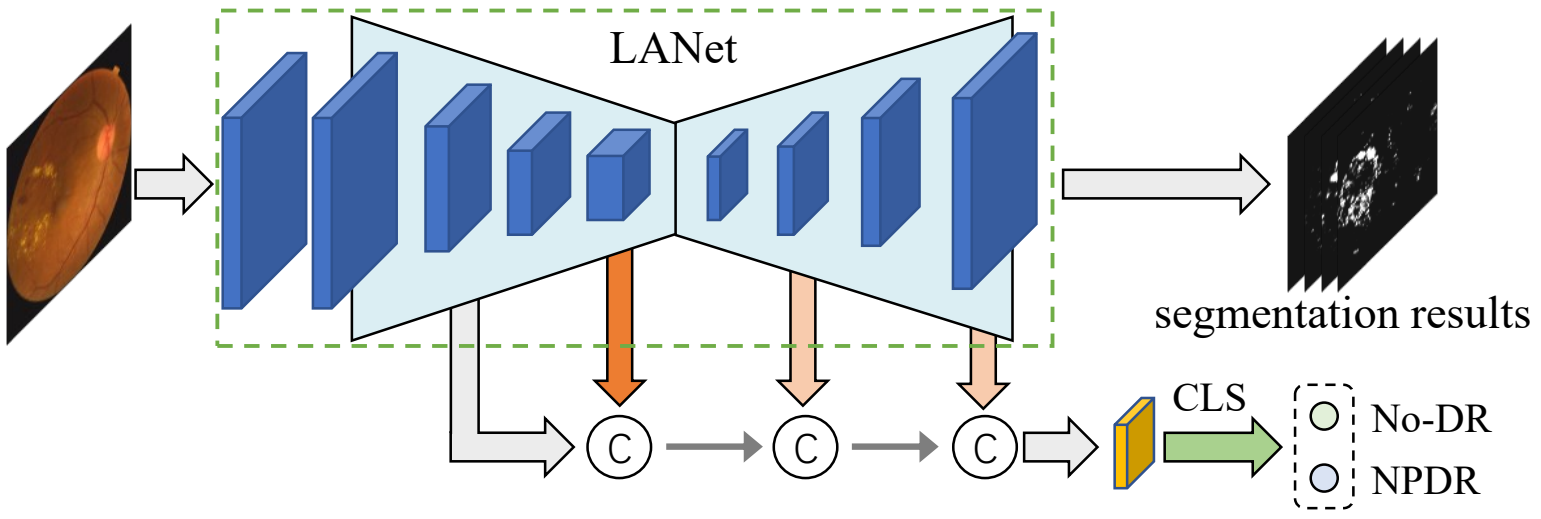

**FIGURE 5** LASNet: Lesion-aware screening network, which is an extension of LANet. *CLS* is short for classification, and circles marked with *C* stand for concatenation along channel.

As shown in Fig. 4, gray arrows indicate convolutional layers while single lines with arrows denote data passing without extra operation. In accordance with the subscripts in Fig. 2, $\boldsymbol{x}_{\text{enc}}$ and $\boldsymbol{x}_{\text{dec}}$ are features from encoding and decoding layers, and $\boldsymbol{x}_{\text{FPB}}$ implicates the attention map computed by FPB. The FFB is explained as follows:

$$\begin{cases} z_1 = f_{\text{conv1}}(\boldsymbol{x}_{\text{enc}}) \\ z_2 = f_{\text{up1}}(f_{\text{conv2}}(\boldsymbol{x}_{\text{dec}})) \\ z_3 = f_{\text{up2}}(\boldsymbol{x}_{\text{dec}}) \end{cases} \tag{4}$$

$$\begin{aligned} \boldsymbol{x}_{\text{fuse}}(i,j,c) = f_{\text{conv3}}(f_{\text{conc}}(& z_1(i,j,c) \times \boldsymbol{x}_{\text{FPB}}(c), \\ & z_2(i,j,c) \times \boldsymbol{x}_{\text{FPB}}(c), \\ & z_3(i,j,c) \times \boldsymbol{x}_{\text{FPB}}(c))) \end{aligned} \tag{5}$$

both of $f_{\text{up1}}(\cdot)$ and $f_{\text{up2}}(\cdot)$ indicate convolutional layers with upsampling step, while $f_{\text{conc}}(\cdot)$ denotes concatenation.

## 3.2 | Lesion-Aware Screening Network (LASNet)

Since our LANet is designed for eventually assisting DR screening, to further adapt the segmentation network to the classification task, we simply add a classification layer to LANet. As shown in Fig. 5, the LANet works as the backbone of the screening network, termed as LASNet (Lesion-Aware Screening Network). The blue hourglass corresponds to the encoder-decoder-structure of the network. The lesion segmentation results are the outputs of LANet, and the classification result is the output of LASNet.

The bold arrows in light gray are operations including convolution, non-linear activation and batch normalization. Those in orange and light orange denote the HAM and the max pooling respectively. While the green arrow denotes the classification layer comprising global average pooling and fully-connected layers. Noting that the LASNet only outputs no-DR and NPDR, as mentioned above, our net conducts screening (binary classification) instead of grading (multi-class classification).

It's clear that there is no extra heavy sub-network when conducting DR screening since the LASNet totally adopted LANet as the backbone. What's more, this design proves that segmentation is able to assist grading as the latter relies heavily on the former.

## 3.3 | Loss Functions

In this paper, the proposed LAM and FPM are stacked and then aggregated into an encoder-decoder structure to construct a lesion segmentation network termed LANet. By adding a classification layer, the LANet can be extended to a DR screening network termed LASNet. The output of LANet is a binary lesion map with four channels, each of which represents one symptoms in HE, MA, EX and SE. While the result of LASNet is a binary scalar indicating No-DR or NPDR.

Therefore, both cross-entropy (CE) and binary cross-entropy (BCE) losses work. In our work, BCE is adopted for pixel-wise segmentation task and CE for screening. However, lesions generally occupies a small part of a whole fundus image, and some samples do not even contain certain types of lesions. To deal with imbalanced distribution of positive pixels (lesion pixels) and negative ones, we involve a weight for positive pixels, formulated as Eq. (6):

$$\ell_{\text{seg}} = -\frac{1}{H \times W} \sum_{i=1}^{H} \sum_{j=1}^{W} \left[\alpha \boldsymbol{g}(i,j) \log \boldsymbol{x}_{\text{out}}(i,j) + (1 - \boldsymbol{g}(i,j)) \log(1 - \boldsymbol{x}_{out}(i,j))\right] \tag{6}$$

where $H$ and $W$ stand for the height and width of an image, $\boldsymbol{g}(i,j)$ means the groundtruth and $\boldsymbol{x}_{out}(i,j)$ means one channel of the predicted map. The $\alpha$ is the the weight for positive pixels that forces network to focus on lesion pixels.

Annotations for medical data do not strictly share the same criterion since annotators may have their own clinical experience [30]. Consequently, there is a possibility of incorrect or inaccurate annotations [31], and over-belief on manually annotated data will lead to over-fitting. To this end, we also apply a label smoothing strategy [32], which has been proved as an effective regularization [33,34], to improve the loss function. As presented in Eq. (7):

$$\ell_{\text{scr}} = -\sum_{i=1}^{N} \hat{y}_i \log\left(\frac{exp(y_i)}{\sum_j exp(y_j)}\right) \tag{7}$$

where $N$ means the sample number, $y_i$ and $\hat{y}_i$ implicate the $i$-th predicted score and its corresponding smoothed label. $\hat{y}_i$ is computed by Eq. (8):

$$\hat{y}_i = \begin{cases} 1 - \varepsilon + \frac{\varepsilon}{C} & if \quad \hat{y}'_i = 1 \\ \frac{\varepsilon}{C} & otherwise \end{cases} \tag{8}$$

where $\varepsilon$ is a hyperparameter controlling smooth level and $\hat{y}'_i$ implicates the original hard label. A larger $\varepsilon$ indicates less trustworthiness while brings more smoothness to labels. In our task, $C = 2$ and $\varepsilon$ was set to 0.2.

## 4 | EXPERIMENTS

In this section, we conducted lesion segmentation, and ablation study to validate the effectiveness of the proposed modules and to present the performance of LANet. In addition, DR screening experiments were conducted to prove that our LANet also works for screening task after some slight modifications.

### 4.1 | Datasets

We tested our method on three public datasets, which are DDR [35], IDRiD [36] and FGADR [37]. Both of them include pixel-level annotations of lesions that support segmentation task. The details of the three datasets are shown in Table 1.

**TABLE 1** Distribution of train, valid and test images per dataset. The dataset used for screening is denoted as “Scr" and those for segmentation are termed with “Seg"

| | Class | Training | Validation | Testing |
|---|---|---|---|---|
| IDRiD-Seg | - | 40 | 14 | 27 |
| DDR-Seg | - | 383 | 149 | 225 |
| FGADR-Seg | - | 920 | 369 | 553 |
| DDR-Scr | No-DR | 3133 | 1253 | 1880 |
| | NPDR | 2671 | 1068 | 1604 |

**TABLE 2** Lesions segmentation results on IDRiD-Seg, DDR-Seg and FGADR-Seg. The best results are bolded

| | MAE | | | | DICE | | | | AP | | | | mAP |
|---|---|---|---|---|---|---|---|---|---|---|---|---|---|
| | EX | HE | MA | SE | EX | HE | MA | SE | EX | HE | MA | SE | |
| **IDRiD-Seg** | | | | | | | | | | | | | |
| DeepLabV3+ [38] | 0.145 | 0.247 | 0.066 | 0.072 | 0.344 | 0.126 | 0.007 | 0.137 | 0.308 | 0.078 | 0.002 | 0.108 | 0.124 |
| UNet [15] | 0.061 | 0.083 | 0.048 | 0.053 | 0.566 | 0.257 | 0.011 | 0.444 | 0.575 | 0.239 | 0.004 | 0.410 | 0.307 |
| HED [39] | 0.022 | 0.043 | 0.009 | 0.017 | 0.557 | 0.337 | 0.127 | 0.236 | 0.566 | 0.294 | 0.058 | 0.221 | 0.285 |
| HED_cGAN [40] | 0.012 | 0.019 | 0.001 | 0.005 | 0.422 | 0.230 | 0.023 | 0.088 | 0.393 | 0.161 | 0.010 | 0.068 | 0.158 |
| EADNet [41] | 0.073 | 0.102 | 0.046 | 0.051 | 0.405 | 0.137 | 0.004 | 0.185 | 0.361 | 0.090 | 0.002 | 0.165 | 0.155 |
| Sambyal et al. [42] | 0.019 | 0.019 | 0.010 | 0.005 | 0.556 | 0.397 | 0.181 | 0.484 | 0.576 | 0.360 | 0.107 | 0.528 | 0.393 |
| MTUNet [18] | 0.033 | 0.056 | 0.023 | 0.031 | **0.626** | 0.435 | 0.099 | 0.337 | **0.684** | 0.419 | 0.052 | 0.296 | 0.363 |
| Swin-B [43] | 0.013 | 0.010 | 0.004 | 0.003 | 0.551 | **0.497** | 0.151 | 0.570 | 0.546 | **0.483** | 0.087 | 0.577 | 0.423 |
| LANet(Ours) | **0.008** | **0.010** | **0.001** | **0.002** | 0.611 | 0.484 | **0.243** | **0.657** | 0.641 | 0.476 | **0.167** | **0.713** | **0.499** |
| **DDR-Seg** | | | | | | | | | | | | | |
| DeepLabV3+ [38] | 0.064 | 0.111 | 0.026 | 0.054 | 0.302 | 0.151 | 0.002 | 0.159 | 0.258 | 0.108 | 0.002 | 0.120 | 0.122 |
| UNet [15] | 0.009 | 0.015 | 0.004 | 0.006 | 0.584 | 0.435 | 0.273 | 0.363 | 0.614 | 0.406 | 0.236 | 0.353 | 0.402 |
| HED [39] | 0.009 | 0.016 | 0.002 | 0.007 | 0.518 | 0.434 | 0.232 | 0.343 | 0.541 | 0.415 | 0.187 | 0.327 | 0.367 |
| HED_cGAN [40] | 0.005 | 0.008 | 4.7E-04 | 0.003 | 0.453 | 0.403 | 0.063 | 0.284 | 0.462 | 0.373 | 0.036 | 0.275 | 0.287 |
| EADNet [41] | 0.025 | 0.042 | 0.008 | 0.016 | 0.447 | 0.245 | 0.032 | 0.212 | 0.431 | 0.184 | 0.022 | 0.171 | 0.202 |
| Sambyal et al. [42] | 0.007 | 0.011 | 0.002 | 0.002 | 0.526 | 0.476 | 0.283 | 0.405 | 0.560 | 0.462 | 0.248 | 0.459 | 0.432 |
| MTUNet [18] | 0.017 | 0.029 | 0.008 | 0.015 | 0.516 | 0.441 | 0.088 | 0.292 | 0.522 | 0.424 | 0.054 | 0.241 | 0.310 |
| Swin-B [43] | 0.009 | 0.017 | 0.010 | 0.003 | 0.540 | 0.492 | 0.257 | **0.503** | 0.562 | 0.487 | 0.204 | **0.575** | 0.457 |
| LANet(Ours) | **0.004** | **0.007** | **4.1E-04** | **0.001** | **0.607** | **0.530** | **0.322** | 0.440 | **0.623** | **0.515** | **0.289** | 0.487 | **0.478** |
| **FGADR-Seg** | | | | | | | | | | | | | |
| DeepLabV3+ [38] | 0.088 | 0.136 | 0.075 | 0.044 | 0.386 | 0.390 | 0.124 | 0.322 | 0.351 | 0.356 | 0.074 | 0.299 | 0.270 |
| UNet [15] | 0.013 | 0.024 | 0.007 | 0.008 | 0.407 | 0.322 | 0.142 | 0.268 | 0.385 | 0.291 | 0.076 | 0.240 | 0.248 |
| HED [39] | 0.033 | 0.055 | 0.028 | 0.020 | 0.401 | 0.351 | 0.163 | 0.161 | 0.383 | 0.310 | 0.098 | 0.116 | 0.227 |
| HED_cGAN [40] | 0.011 | 0.017 | 0.006 | 0.006 | 0.497 | 0.463 | 0.159 | 0.324 | 0.517 | 0.463 | 0.130 | 0.292 | 0.351 |
| EADNet [41] | 0.059 | 0.096 | 0.042 | 0.037 | 0.432 | 0.423 | 0.144 | 0.380 | 0.443 | 0.407 | 0.095 | 0.362 | 0.327 |
| Sambyal et al. [42] | 0.018 | 0.028 | 0.010 | 0.007 | 0.016 | 0.471 | 0.469 | 0.169 | 0.496 | 0.472 | 0.131 | 0.432 | 0.383 |
| MTUNet [18] | 0.019 | 0.029 | 0.017 | 0.013 | 0.504 | 0.493 | 0.224 | 0.371 | 0.521 | 0.499 | 0.149 | 0.353 | 0.381 |
| Swin-B [43] | 0.024 | 0.044 | 0.026 | 0.014 | 0.484 | 0.499 | 0.189 | 0.411 | 0.527 | 0.517 | 0.133 | **0.483** | 0.415 |
| LANet(Ours) | **0.008** | **0.011** | **0.005** | **0.004** | **0.532** | **0.519** | **0.259** | **0.424** | **0.532** | **0.518** | **0.198** | 0.445 | **0.423** |

- IDRiD consists of 81 fundus images annotated with 4 types of DR-related lesions, *i.e.*, HE, MA, EX and SE. This dataset is involved to demonstrate the performance of LANet, and is termed as IDRiD-Seg, where *Seg* is short for segmentation.
- DDR contains 13673 fundus images with image-level labels of 6 classes, including No-DR (0), Mild (1), Moderate (2), Severe (3), Proliferative (4) and Ungradable (5). The grading labels are determined according to the ICDR. As mentioned in introduction, we only involve data of No-DR (0) and NPDR ($1 \sim 3$) in our screening task. Thus there are 11609 images for DR screening, among which 757 images are utilized for segmentation. We termed them as DDR-Scr and DDR-Seg respectively, where *Scr* is short for screening.
- FGADR provides 1842 images with grading labels and pixel-level lesion annotations as well. The lesions include MA, HE, EX, SE, and other two. To conduct a fair comparison, we only adopted the former four lesions, discarded 287 PDR images and randomly split a validation set. The grading annotation also follows the $0 \sim 4$-level criterion of ICDR.

## 4.2 | Implementation Details

In lesion segmentation, the LANet was trained on DDR-Seg, IDRiD-Seg and FGADR, respectively. The backbone of LANet is ResNet-50 pre-trained on ImageNet [44]. All inputs are pre-processed RGB fundus images in size of $512 \times 512$. Besides, black areas cropping and adaptive histogram equalization were conducted to enhance the quality of images. Also, random horizontal flipping, random rotation and random crop were adopted as data augmentation. The batch size was set to 8, and SGD optimizer with dynamic learning rate was utilized.

While in DR screening, the trained LANet was leveraged as a pre-trained model since LASNet for screening share the most modules with LANet. Afterwards, the LASNet was fine-tuned on DDR-Scr. All inputs were applied with the same pre-processing as those in LANet. The batch size was remained as 8, and AdamW optimizer [45] with cosine annealing learning rate was utilized.

All experiments are conducted on a computer with an AMD Ryzen 5950X Processor, 64GB RAM and an Nvidia GeForce RTX 3090 GPU.

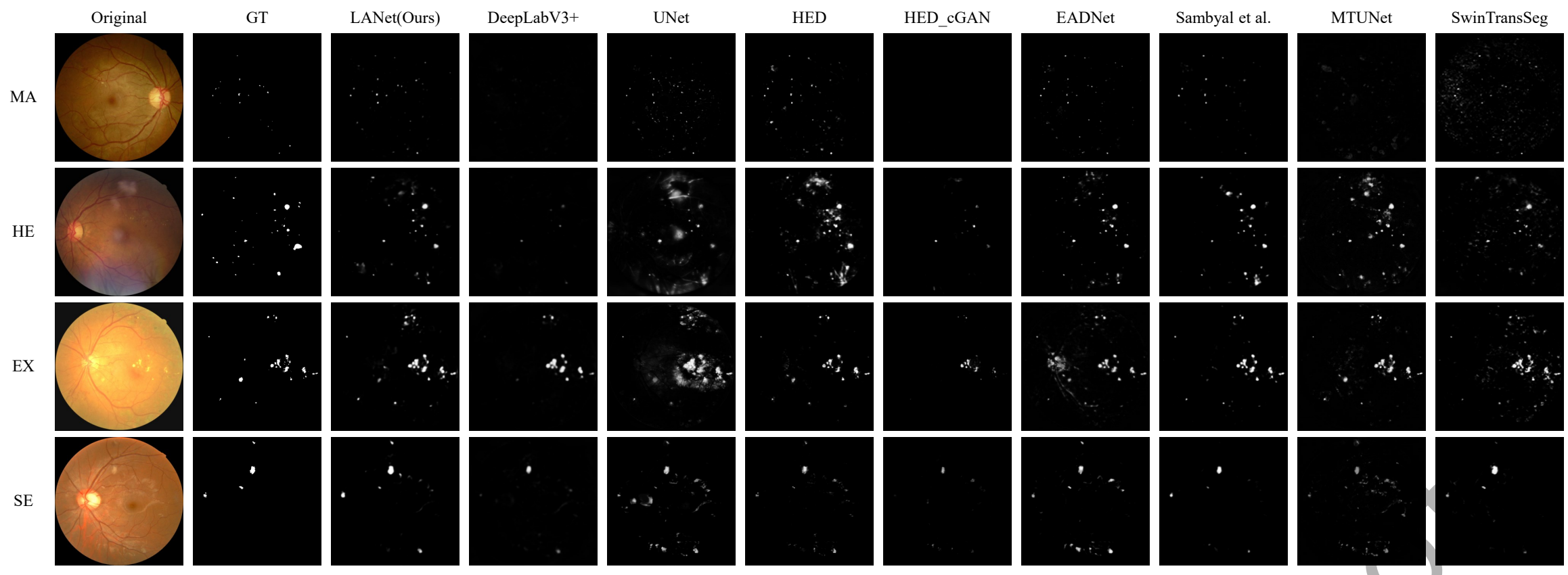

**FIGURE 6** Visualization of lesion segmentation results on a testing sample from DDR-Seg.

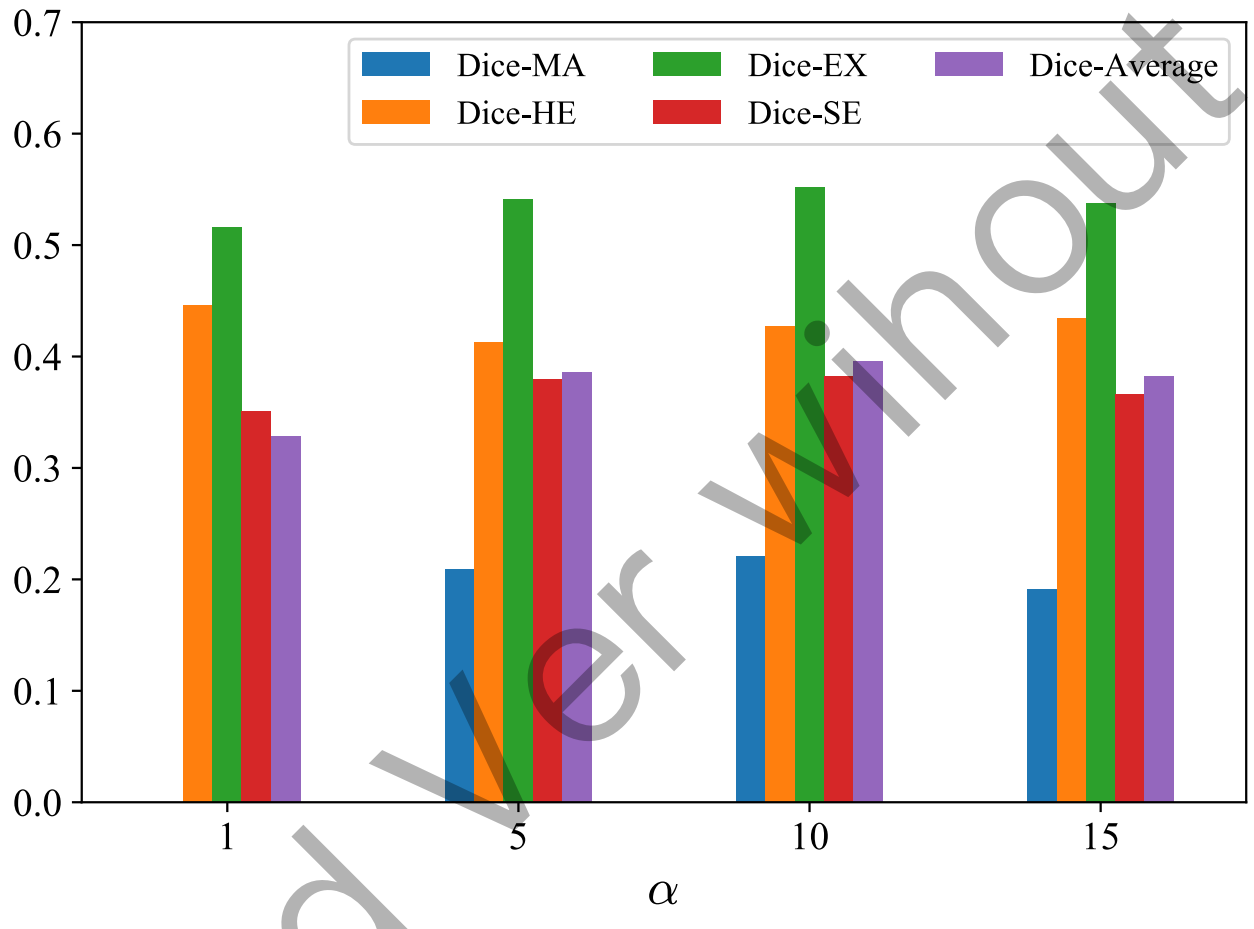

**FIGURE 7** The segmentation results in the term of Dice score with different $\alpha$ values.

## 4.3 | Evaluation on DR lesion segmentation

**Metrics.** Mean Absolute Error (MAE), Dice score, Average Precision (AP) and mean of AP (mAP) were adopted as metrics. Noting that MAE, Dice score and AP were computed within a single type of lesion while mAP is computed over all samples across lesions. As illustrated below:

$$AP = \sum_{m}(R^m - R^{m-1})P^m \tag{9}$$

$$mAP = \frac{1}{K}\sum_{k=1}^{K} AP_k \tag{10}$$

where $R^m$ and $P^m$ are recall and precision at the $m$th threshold computed from a specific type of samples. $K$ is the number of lesion types, and $AP_k$ is the AP of the $k$th lesion.

$$Dice = \frac{1}{N}\sum_{i}^{N}\frac{2(\boldsymbol{y}_i \cap \hat{\boldsymbol{y}}_i)}{\boldsymbol{y}_i + \hat{\boldsymbol{y}}_i} \tag{11}$$

$N$ is the sample number. $\boldsymbol{y}_i$ and $\hat{\boldsymbol{y}}_i$ implicate the predicted lesion map of the $i$-th fundus image and its corresponding groundtruth map.

**Results Analysis.** The experimental results of DR lesion segmentation are shown in Table 2. We adopted *exactly the same* pre-processing and hyper-parameters in both DeepLab V3+ (Xception as backbone) [38], UNet [15] ,HED [39], HED_cGAN [40] , EADNet [41] Sambyal et al.[42], MTUNet [18] and Swin-B [43] in all comparisons.

Our LANet obtains the best performance on both DDR-Seg and IDRiD-Seg overall, espacially in terms of MA, which is a tiny symptom. This proves that our LANet is able to capture small lesion from data with extremely imbalanced pixel distribution.

On IDRiD-Seg dataset, MTUNet and Swin-B beat us in identifying EX and HE respectively. The former lesions appear as bright dots with hard edges and the latter exhibit irregular blood areas. MTUNet with relatively less layers is able to detect visible regular areas, like EX, through convolutions. While identifying irregular areas with blur edges depends on not only local information but also long range features. Therefore, the global self-attention modules in Swin Transformer successfully recognize HE. Overall, we still achieve the best mAP and MAE.

Fig. 6 demonstrates the visualization of some lesion segmentation results on DDR-Seg. It's clear that DeepLab V3+ fails in capturing tiny lesion areas, the possible reason is that the ASPP module possesses too big receptive fields for MA and HE. UNet is not robust to light areas or noise since it was originally designed for cell segmentation, which is a task that focus on areas with more visible gradient differences. EADNet also achieves good results since it involves dual attention [12] in mid-layers to obtain lesion information along spatial position and channel. While our attention modules are specifically designed and are applied to the whole decoder stage, thus we suppress some false positives. HED and HED-cGAN were trained on individual lesion, thus we modified them to a multi-lesion segmentator and re-trained it for fairness. As a results, they suffer from false positive or false negative when trained for simultaneous 4-lesion-segmentation. In all, our LANet outperforms others despite several false alarmed SE dots.

Fig. 7 shows segmentation results in the term of Dice score with a varied $\alpha$. As mentioned in Eq. (6), $\alpha$ represents the weight of positive pixels, and it was varied from 1 to 15 in a stride of 5. It's obvious that assigning proper bias (*i.e.*, $\alpha$) helps the network balance the focus on lesion pixels. According to Fig. 7, $\alpha$ was set to 10 in our experiments.

## 4.4 | Ablation Study

To gain insights of the proposed network, we conducted ablation by removing LAM and FPM. To successfully run the ablated network, we made the following modification to LANet in the ablation experiment.

- For fair comparison, we replaced the LAM with a covolution layer and replaced FPM with bilinear interpolation to build a base version of the comparison network, termed as Base.
- To validate the effectiveness of LAM, we added LAM to Base network, termed as Base+LAM.
- To validate the effectiveness of FPM, we involved FPM in Base network, termed as Base+FPM.
- At last, we added both LAM and FPM to construct the proposed LANet, termed as Base+LAM+FPM.

The results of ablation are shown in Table 3. It's clear that both modules works. FPM boosts the performance, especially in detecting tiny lesions, *i.e.*, MA. The reason may be that the FPB and FFB in FPM involve both low-level features with a smaller receptive field indicating lesions and global attention implicating diseases.

Although Base+LAM obtains lower scores on SE, the Base+LAM+FPM achieves the best performance on all lesions. This indicates that LAM works better when combined with FPM than being involved alone. The reason maybe that LAM contributes to small lesion description such as HE, since the orientation-aware sub-module in LAM captures gradients in a relatively small neighbors. However, the information captured by LAM is not well leveraged in Base+LAM network. While FPM leverages the information in a better way than simply concatenation or summation: 1) FPM adopts local lesion-related and global disease-related information, 2) LAM+FPM is stacked as a multi-layer decoder rather than being used only once as a single layer.

**TABLE 3** Ablation results for lesion segmentation in the term of Dice score of each lesion. The best results are bolded.

| | EX | HE | MA | SE |
|---|---|---|---|---|
| Base | 0.177 | 0.304 | 0.001 | 0.243 |
| Base+LAM | 0.215 | 0.323 | 0.001 | 0.093 |
| Base+FPM | 0.582 | 0.525 | 0.277 | 0.416 |
| Base+LAM+FPM | **0.607** | **0.530** | **0.322** | **0.440** |

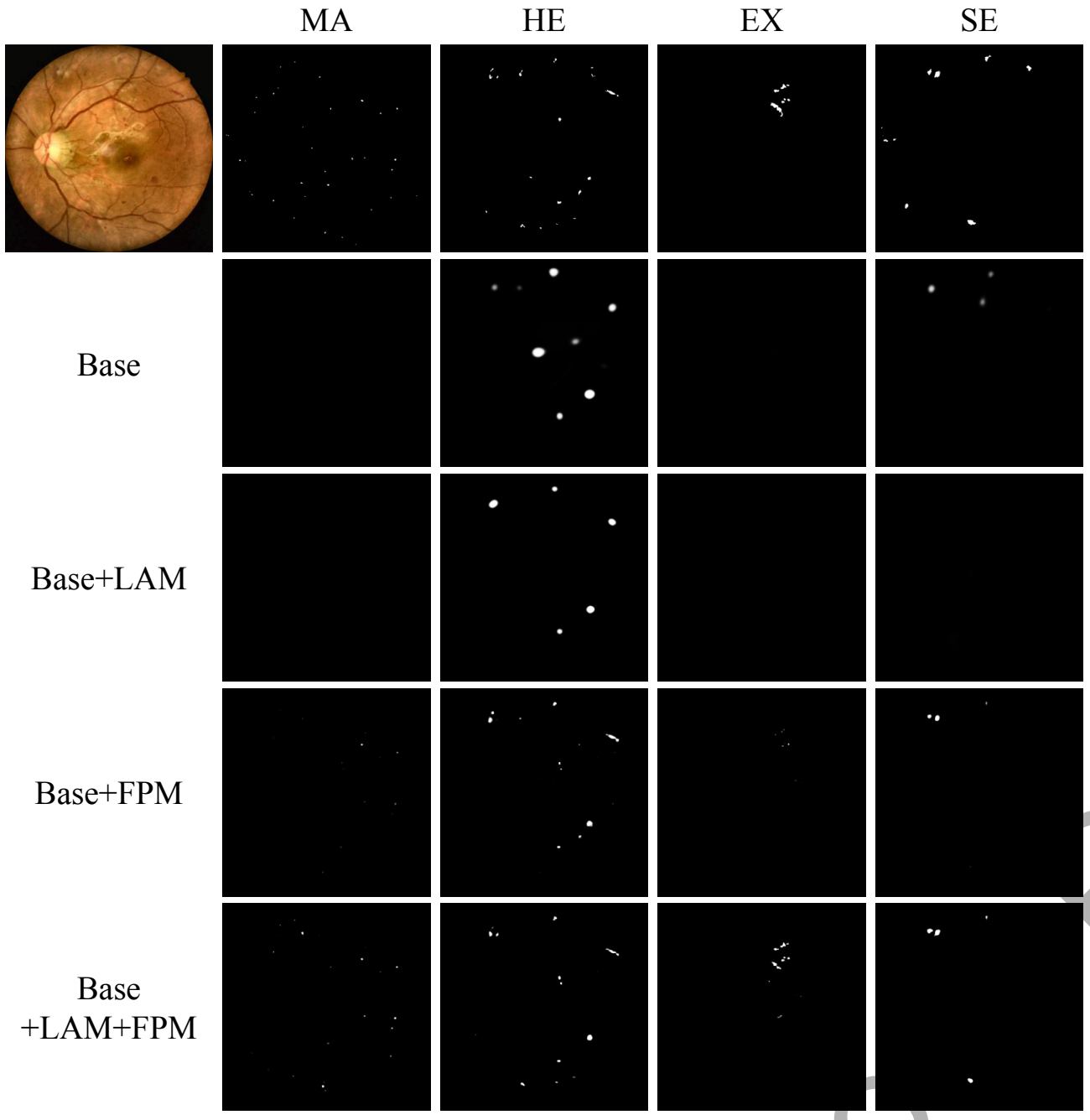

**FIGURE 8** Visualization of ablation study.

This can be verified by Fig. 8 , in which the first row locate the groundtruths. It's obvious that Base net obtains false positives on HE and false negatives on other three lesions, and some of them are corrected by LAM and FPM individually. While the LAM+FPM successfully helps recognize other small lesions, especially the extremely tiny MA. Fig. 9 illustrates the output of the 4 decoding layers, where (a)~(d) respectively stands for the masks of $\boldsymbol{x}_{dec}^1 \sim \boldsymbol{x}_{dec}^4$ and groundtruths lie in the first row. It is clear that the outputs become more accurate when decoder goes deeper. Thus, we conclude that the LANet explores lesion-aware features progressively across layers.

## 4.5 | Evaluation on DR Screening

**Metrics.** Precision, Sensitivity, also known as True Positive Rate (TPR), F1-score, and Area Under Curve(AUC) were adopted as metrics.

**Results Analysis.** In order to prove that our LANet not only performs accurate segmentation but also benefits the DR screening, we extended LANet to LASNet by simply adding a classification layer without any extra trainable modules. Two versions of training were involved in DR screening as a comparison:

- LASNet was trained from scratch for 30 epochs.
- LASNet was fine-tuned on DDR-Scr for 20 epochs with a fixed learning rate of $3 \times 10^{-4}$ based on the trained LANet.

The curves in Fig. 10 show that the pre-trained LASNet converges faster and obtains higher accuracy on validating data. This proves that a well-trained segmentation model can greatly help DR screening, since the segmentation and screening networks share the same trainable structure except for an extra output layer.

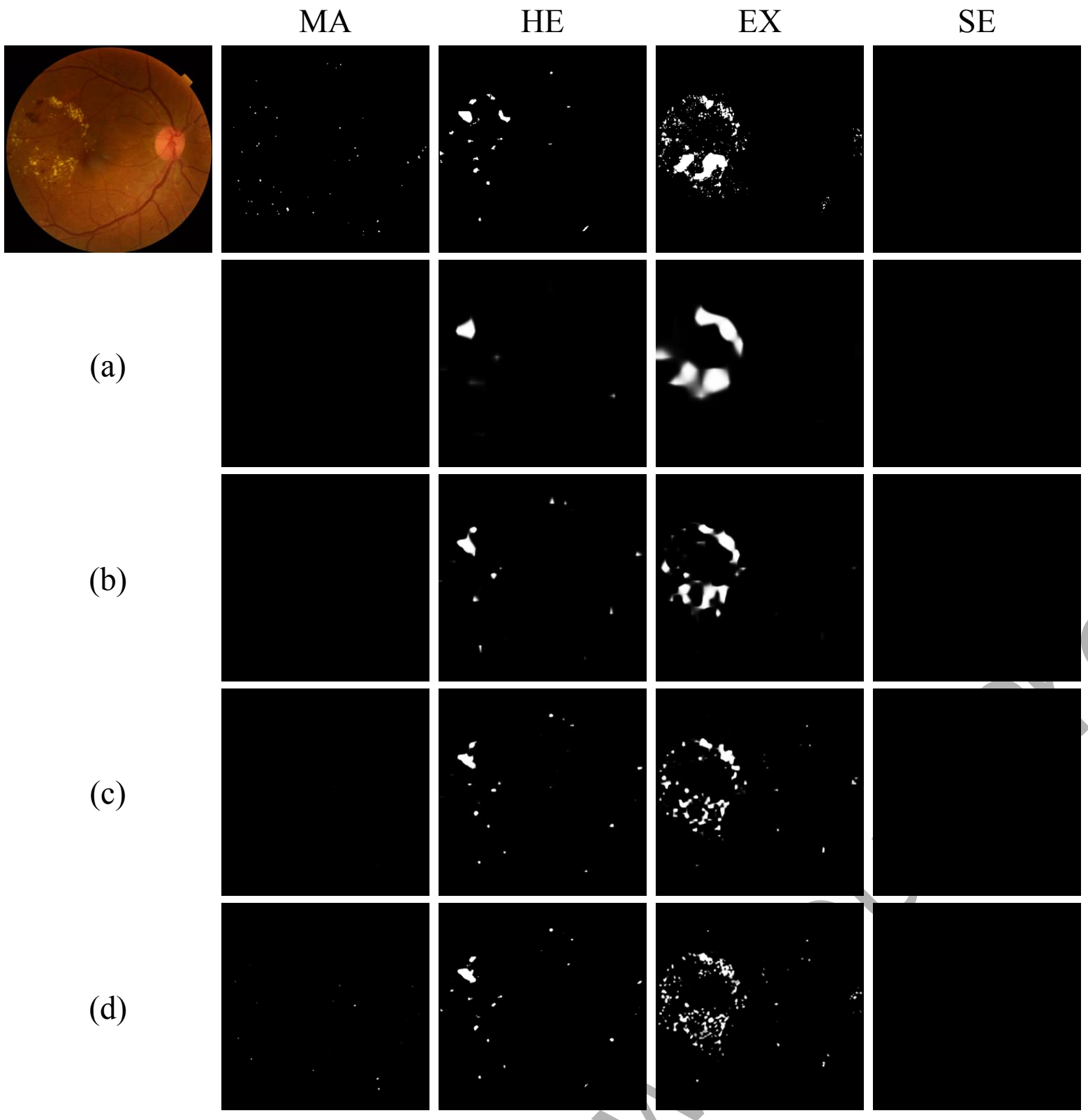

**FIGURE 9** Visualization of the output from different decoder layers. The groundtruths locate in the first row, and the masks in (a)~(d) represent the outputs of 1~4th decoder layer.

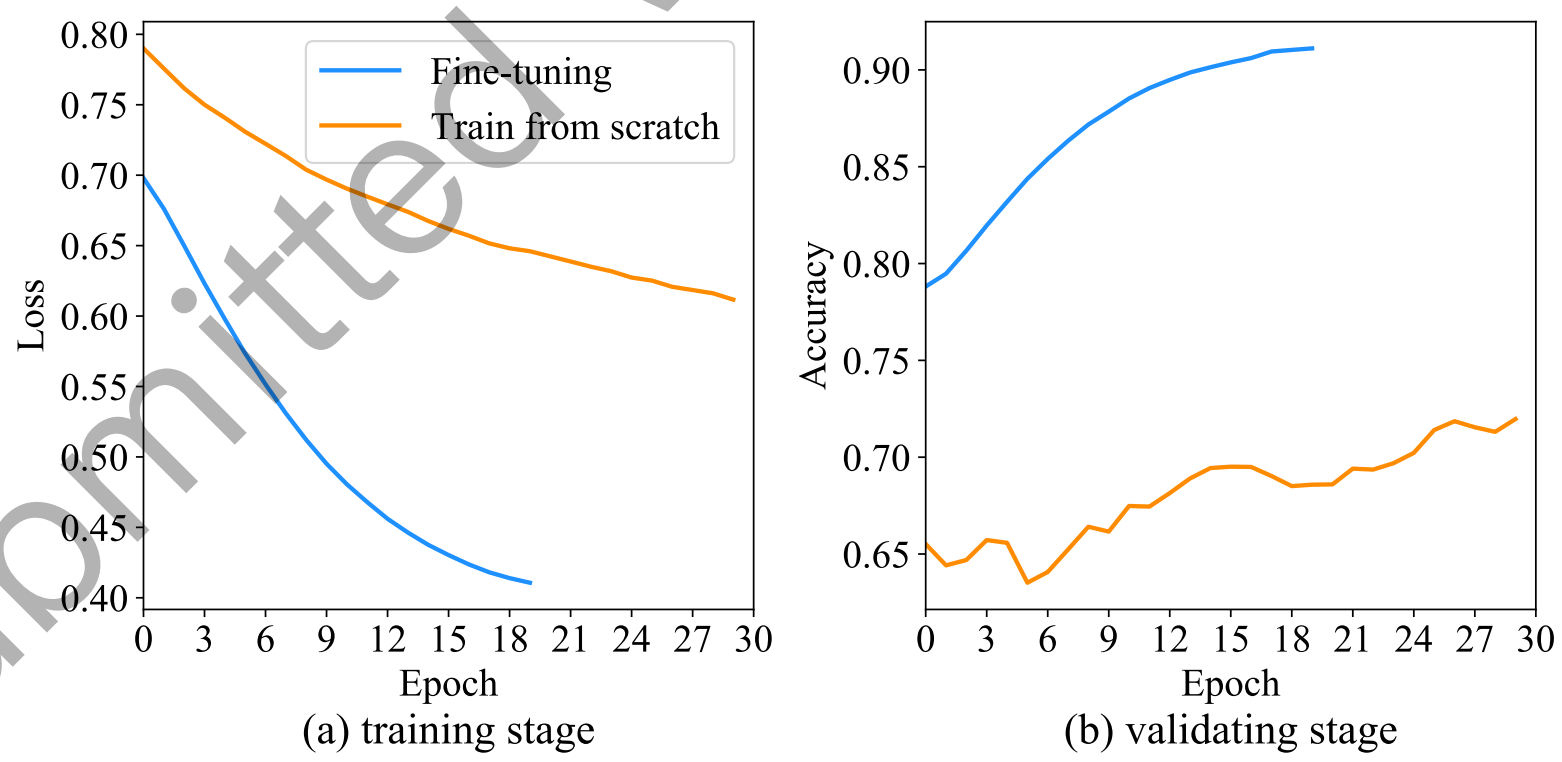

**FIGURE 10** Comparison between the LASNet trained from scratch and the one pre-trained on DDR-Seg.

Furthermore, we involved ResNet-34[27], ResNet-50[27], Inception V3[32], MTUNet[18] and CABNet[14] as comparison methods for DR screening. In CAB model, the hyper-parameter $k = 2$. Comparison metrics are recorded in Table 4. Noting that ResNet-18 and ResNet-50 obtain similar results, the reason may be that some tiny lesions are progressively ignored or mis-classified as network goes deeper. While our LASNet preserves these features by short connection with FPM and LAM.

Although the our Pr is 1.1% lower the best records, the Pr (Precision) only quantifies the number of DR images that actually belong to DR class at a specific threshold. While F1 calculates the harmonic mean of Pr and Se, which balances both the concerns of Pr and Se. Similarly, AUC also indicates the performance of LASNet at distinguishing between DR images and non-ones.

**TABLE 4** DR screening results on DDR-Scr. Acc: Accuracy, Pr: Precision, Se: Sensitivity.The best results are bolded.

| | Pr | Se | F1 | AUC |
|---|---|---|---|---|
| **DDR-Scr** | | | | |
| ResNet-34[27] | **0.973** | 0.736 | 0.868 | 0.954 |
| ResNet-50[27] | 0.958 | 0.774 | 0.879 | 0.954 |
| Inception v3[32] | 0.958 | 0.791 | 0.888 | 0.950 |
| MTUNet[18] | 0.875 | 0.461 | 0.722 | 0.832 |
| CABNet[14] | 0.970 | 0.746 | 0.873 | 0.953 |
| LASNet(Ours) | 0.962 | **0.840** | **0.911** | **0.967** |

Therefore, F1 and AUC offers more general evaluation on diagnosing performance, and our F1 and AUC is 2.3% and 1.3% higer than the 2nd-ranked records. Overall, we achieves improvements over other comparison networks.

## 5 | CONCLUSION

In this paper, we proposed an LANet for DR diagnosis, specifically for DR-related lesion segmentation by embedding lesion-aware module (LAM) and feature-preserve module (FPM). The former module aims at progressively exploring lesions through attention, while the latter learns to leverage shallow-to-deep features by preserving lesion-related local information and disease-related global features. Through constructing a classification layer consisting of global average pooling and fully-connected layers, the LANet can be easily extended to DR screening task, denoted as LASNet. Experiments prove that our LANet outperforms other methods in capturing tiny lesions by acquiring and preserving lesion-aware features. Moreover, the ablation study validates that the combination of LAM and FPM designed for segmentation also benefits the DR screening.

Since it has been proved that global and local attention benefits irregular and tiny lesions respectively, we will further exploring these attentions for better HE and SE recognition. Specifically, our future work will mainly focus on 1) involving global attention for exploring irregular DR-related lesions, 2) designing structures to make better use of cross-layer-information that works for tiny lesions and ambiguous areas.

## 6 | ACKNOWLEDGEMENTS

This work is partly supported by the National Natural Science Foundation 62162029 and Natural Science Foundation of Jiangxi Province 20202BABL212007.

## 7 | ORCID

Xue Xia: 0000-0002-2872-7151
Kun Zhan: 0000-0002-0614-3489
Yuming Fang: 0000-0002-6946-3586
Wenhui Jiang: 0000-0002-4144-6725
Fei Shen: 0000-0001-9885-3316